\documentstyle[twocolumn,aps]{revtex}
\input epsf
\draft

\begin{document}

\twocolumn[\hsize\textwidth\columnwidth\hsize\csname@twocolumnfalse\endcsname

\title{Linear and Nonlinear Experimental Regimes of Stochastic Resonance}
\author{Rosario N. Mantegna$^{*}$, Bernardo Spagnolo$^{*}$ 
and Marco Trapanese$^{\#}$}
\address{$^{*}$INFM, Unit\'a di Palermo, and Dipartimento di 
Fisica e Tecnologie Relative, Universit\'a di Palermo,\\ 
Viale delle Scienze, I-90128 Palermo, Italia\\ 
$^{\#}$ INFM, Unit\'a di Palermo, and Dipartimento di
Ingegneria Elettrica, Universit\'a di Palermo,\\
Viale delle Scienze, I-90128 Palermo, Italia\\}

\date{\today}

\maketitle

\begin{abstract}

We investigate the stochastic resonance phenomenon in a physical
system based on a tunnel diode. The experimental control parameters 
are set to allow the control of the frequency and amplitude of the 
deterministic modulating signal over an interval of values
spanning several orders of magnitude. We observe both a regime
described by the linear response theory and the nonlinear deviation
from it. In the nonlinear regime we detect saturation of the power 
spectral density of the output signal detected at the frequency 
of the modulating signal and a dip in the noise level of the same
spectral density. When these effects are observed we detect a
phase and frequency synchronization between the stochastic output 
and the deterministic input.  

\end{abstract}

\pacs{05.40.-a, 85.30.Mn}
\vskip1pc]

\narrowtext

\section{Introduction.}

The renewed interest of the last two decades on
stochastic processes modeling different phenomena
of physics, chemistry and engineering sciences has 
led to the discovery of noise-induced phenomena 
in nonlinear systems away from equilibrium. 
In these systems a variation of the
level of external noise can qualitatively
change the response of the system. 
The paradigmatic example
of these noise-induced phenomena is stochastic resonance
\cite{Vulpiani} (for a recent review see \cite{Gammaitoni98}).
Other examples of noise-induced phenomena comprise resonant 
activation \cite{Doering}, noise-induced transitions 
\cite{VandenBroeck} and noise enhanced stability \cite{Weiss,MS96}.

Stochastic resonance (SR) manifests itself as an enhancement of the
system response for certain finite values of the noise strength. In
particular the signal-to-noise ratio (SNR) shows a maximum as a
function of the noise intensity. In other words, a statistical
synchronization of the random transitions between the two
metastable states of the nonlinear system takes place in
the presence of an external
weak periodic force and noise. Such a physical system presents
a time-scale matching condition, which can be observed 
by tuning the noise level to such a value that the period of the
driving force approximately equals twice the noise-induced
escape time. The SR phenomenon appears in a large variety of
physical systems and has been observed in different systems,
ranging from sets of neurons \cite{WiesenfeldMoss}, to lasers 
\cite{Roy} and to solid-state devices, like SQUIDs and tunnel 
diodes \cite{MS94}.

The SR phenomenon is a well investigated phenomenon both 
from a theoretical and an experimental point of view
\cite{Gammaitoni98}. However, few studies systematically 
analyze the SR phenomenon for different values 
of the frequency and amplitude of the modulating signal
\cite{Roy,MS94,McWiesenfeld,MS95,Giacomelli}.
In this article we systematically study the SR phenomenon
as a function of the frequency and amplitude of the modulating
signal in a physical bistable system based on a tunnel 
diode. Our experimental set-up allows us to investigate this 
phenomenon in a range of amplitude and frequency spanning
several orders of magnitude. By varying the amplitude and the
frequency of the modulating signal we detect both the regime
of the SR phenomenon described by the linear response theory 
and the nonlinear deviation from it. In the linear regime we 
observe the customary behavior of stochastic resonance 
whereas in the nonlinear regime we detect a saturation 
of the power spectral density measured at the frequency of 
the modulating signal and a depletion of the power spectral 
density of the noise at the same frequency. When the noise 
depletion takes place we observe phase and frequency 
synchronization between the stochastic output  and the 
deterministic input.

The paper is organized as follows. In section II we describe the
experimental apparatus and we discuss the stochastic differential
equation associated to the electronic circuit based on a tunnel 
diode. In section III we present our experimental results of 
the power amplification and SNR as a function of the noise 
intensity for different values of the amplitude and frequency of the 
modulating signal. In this section we discuss the detected deviations 
form the predictions of the linear response theory and we 
present an evidence of phase and frequency synchronization 
detected in the nonlinear regime of high values of the 
amplitude of the modulating signal. In section IV we briefly 
draw our conclusions.

\section{Experimental apparatus and the tunnel
diode}

The experimental setup used for investigating the SR phenomenon
is a bistable electronic system based on a tunnel diode.
The physical system is a series of a resistor(tunable to a 
desired value) and a tunnel diode in parallel to a capacitor.
The tunnel diode is a highly doped semiconductor device with
a typical current-voltage characteristic showing a region of
negative differential resistance, which is due to
a tunneling current from the valence band of the n-doped region 
to the conducting band of the p-doped region.
There are two stable states and one unstable state 
\cite{Landauer62,Land78,HanggiT82}. For details about this 
experimental set-up see ref. \cite{LMSZ97}.

A network of general purpose very low-noise wide band
operational amplifier is used to sum
the driving periodic signal and the noise signal.
The noise signal is the output of a commercial digital noise source,
whose spectral density is approximately flat up to 20 MHz
and whose root mean square voltage $V_{rms}$ may be
selected within the range from $0.133$ to $5.5$ V with a 27 mV
resolution. 
At the output of the operational amplifier the statistical
properties of the noise are altered by the filtering of
the operational amplifier. We measure the noise $v_n(t)$
at the output of the operational amplifier and we observe that
it is a Gaussian noise characterized by a spectral density 
which is flat at low frequency ($f<1.25$ MHz).
By defining the correlation time of 
the Gaussian noise as the time at which the normalized 
autocorrelation function assumes the value $1/e$, 
we measure $\tau_n=120.$ ns.
In our measurements we vary the amplitude and 
the frequency of the driving periodic signal 
$v_s(t) = V_{s}cos(2\pi f_s t)$.
Specifically, we vary the amplitude $V_s$ from $0.0067$ to
$1.00$ V, and the frequency $f_s$ from $1$ Hz to $1$ MHz.
The output voltage across the diode $v_d$ is detected by a
digital oscilloscope and transferred to a PC.
A typical time series has 4096 records. The digitized time 
series are analyzed on-line by using a fast 
Fourier transform (FFT) routine. The result of
the FFT routine provides the power spectral density 
of the $v_d$ time signal.

We model our electronic circuit by writing down its
differential equation. This equation is
\begin{equation} 
\frac{dv_{d}}{dt} ~=~
-\frac{dU(v_{d},t)}{dv_{d}} + \frac{1}{RC}v_n(t),
\label{diode-langevin eq}
\end{equation}
which is formally equivalent to a stochastic differential equation
describing the position of an overdamped random particle 
moving in a generalized potential. In this equation $R$ is 
the biasing resistor, $C$ is the parallel capacitor (in our case 
45 pF) of the circuit and $v_n(t)$ is a noise voltage mimicking 
the presence of a finite temperature in the corresponding 
overdamped system with a physical particle. 
The generalized potential $U(v_d,t)$ associated to our physical
system is \cite{LMSZ97}
\begin{eqnarray} 
U(v_d,t)=-\frac{V_b v_d}{RC}+\frac{v_d^2}{2 R C}+
 \frac{1}{C} \int_0^{v_d}  I(v) dv \nonumber \\ 
 -\frac{V_s v_d}{RC} \sin(\omega_s t),
\label{potential}
\end{eqnarray}
here $V_B$ is the biasing voltage of the electronic bistable 
network.

Our circuit presents two control parameters affecting
the shape of the associated generalized potential. They are
the biasing resistance $R$ and the biasing voltage $V_b$. 
We control both of them independently.
We perform our experiments by ensuring a symmetric escape from one 
potential well to the other. This is done by selecting the values 
of the two control parameters (R=770 $\Omega$ and $V_b=6.76$ V)
in such a way that we do not detect power spectral density of the
output voltage $v_d$ above the noise level 
at even harmonics of the frequency of modulating signal. 
We also verify that for this choice of 
control parameters the experimentally measured residence times 
have approximately the same value in the two potential wells. 

The aim of this study is to investigate the SR phenomenon 
over a wide interval of the frequency $f_s$ of the 
modulating signal. To perform such a task, we have to
overcome two experimental conflicting constraints:
(i) we are forced to set the time constant of our system 
$\tau \equiv RC$ to a low value satisfying the 
inequality $f_{s} \ll 1/ 2 \pi \tau$ and (ii) we need to 
use a high value of $\tau$ to maintain the ratio 
$\tau_n/\tau$ as low as possible to conduct our experiments
in the``white noise" limit of $v_n(t)$. The best compromise
we find is to set $\tau \equiv R C=34.6$ ns. With this choice 
$1/ 2 \pi \tau \approx 4 f_s^{max}$ and 
$\tau_n/\tau \approx 3.47$.
In other words we guarantee the investigation of the SR
phenomenon over a rather wide range of $f_s$ by performing
our experiments in a regime of moderately colored noise.

\section{Stochastic Resonance for different values of frequency 
and amplitude of the modulating signal}

A bistable system based on a tunnel diode provides a
versatile physical system in which stochastic resonance can
be investigated \cite{MS94}. In this study we perform 
an investigation of the SR phenomenon as a function
of a wide range of amplitude and frequency of the modulating 
signal. The first investigation concerns the power $P_1$ of the 
output signal $v_d(t)$ localized around the frequency of 
the modulating signal. This quantity is obtained by 
integrating the spectral density $S(\omega)$ over the 
delta-like peak observed at angular frequency $2\pi f_s$.
The signal ``power" $P_1$ used in the theory of linear signal 
processing obtained by integrating the spectral density over the
$f$ peaks at $f_s$ and -$f_s$ is 
\begin{equation}
P_1=4 \pi |M_1|^2,
\end{equation}
here $|M_1|$ is the magnitude of coefficient of the Fourier 
series $<v_d(t)>=\sum_{-\infty}^\infty M_n \exp (i n 2 \pi f_s t)$ 
taken at the frequency of the modulating signal.
The linear response theory for stochastic resonance predicts that
$|M_1|$ is proportional to $V_s$ at fixed value of $V_n$ \cite{Jung91}.
We test this dependence on a large interval of values of the
amplitude signal $V_s$. In Fig. 1 we show the measured values of $|M_1|$
as a function of $V_s$ varying in the interval from 0.0067 to 1.00 V.
The measure are done by setting $f_s=10$ Hz and $V_{rms}=1.89$ V.
From the figure is evident that the prediction $|M_1| \propto V_s$ obtained
by using the linear response theory is valid only within 
the amplitude interval $0.017<V_s<0.067$ V. The deviation observed 
for the lowest investigated value of the modulating amplitude 
signal ($V_s=0.0067$ V) is probably due to experimental detection 
problems related with the low value of the signal whereas the 
deviation observed when $V_s>0.067$ is entirely ascribed to a 
deviation of the physical system from the behavior predicted by the linear
response theory. In particular a saturation of the power 
localized at $f_s$ is detected for large values of the 
amplitude of the modulating signal.
    
The second investigation concerns the frequency dependence
of $|M_1|$. Within the framework of the linear
response theory, at fixed values of $V_n$, $|M_1|$ is 
related to $V_s$, $f_s$ and $V_n$ through the relation 
\begin{equation}
|M_1| \propto <v_d^2> \frac{V_s}{V_{rms}^2} \frac{\lambda_{min}}
{(\lambda_{min}^2+(2\pi f_s)^2)^{1/2}},
\end{equation}
where $<v_d^2>$ denotes a stationary mean value of the unperturbed
system and $\lambda_{min}$ is the smallest non-vanishing eigenvalue of 
the Fokker-Plank operator of the system without periodic driving
\cite{Jung91}. This quantity is an exponential function of the 
noise amplitude $V_{rms}$  under the hypothesis 
of white noise. We set $V_s$=0.067 V to ensure that we are 
in a region of parameters where the linear response theory may 
apply and we perform our experiments as a function of $f_s$ 
for various values of $V_{rms}$. In Fig. 2 we show the results
obtained. The general trend predicted by Eq. (4) is observed.
When $f_s<<\lambda_{min}/2\pi$ a constant value of $|M_1|$ 
is detected whereas in the opposite regime the value of $|M_1|$ 
is decreasing as a function of the frequency. Concerning the 
functional form of $|M_1|$, we observe that 
$|M_1| \propto f_s^{-1.3}$. This is close but not coincident 
with the behavior expected from the linear response theory 
$|M_1| \propto f_s^{-1}$. The observed deviation might be 
ascribed to one or more than one of the following 
possibilities: (i) a distortion introduced by the noise 
background present in our measurements; (ii) an
additional frequency dependence which is present through the term
$<v_d^2>$ of Eq. (4)and (iii) the colored nature of the noise 
$v_n (t)$.

One key aspect of the SR phenomenon is the statistical 
synchronization that takes place when the Kramers time
$T_K(V_{rms})$ between two noise induced inter-well transition is of the 
order of half period of the periodic forcing. In other words 
statistical synchronization occurs when $f_s=1/2 T_K(V_{rms})$.
In our measurement, we verify the validity of this description 
with the following procedure. We set $V_s=0.067$ V and we 
measure the SNR of the output signal $v_d(t)$ at the frequency 
of the modulating signal for six values of $f_s$ ranging 
from 1 to 10$^5$ Hz. We use these experimental results 
to single out for which value of $V_{rms} \equiv V^*_{rms}(f_s)$ 
a maximum of the SNR is detected. This is of course the state 
of maximal statistical 
noise induced synchronization. We then compare $V^*_{rms}(f_s)$
with the function $y(V_{rms})=r_K(V_{rms})/2$, where the Kramers 
rate $r_K(V_{rms})$ is measured in the absence of
a modulating signal. The results are shown in Fig. 3. 
From the figure it is clear that statistical synchronization is observed for
all the investigated frequencies, supporting the traditional 
interpretation \cite{Vulpiani,Gammaitoni98} of the stochastic resonance 
mechanism over a frequency range of the modulating signal spanning 
six frequency orders of magnitude. It is worth pointing out that
synchronization is observed in spite  of the fact that our
experiments are performed in a regime of moderately colored noise.
     
We now investigate the SR phenomenon by studying both 
the signal power amplification 
\begin{equation}
\eta=\frac{P_1}{P_{in}}=4\left[\frac{|M_1|}{V_s}\right]^2
\end{equation}
and the signal to noise ratio 
\begin{equation}
SNR=10 \log_{10}\left[\frac{P_1}{N_1}\right].
\end{equation}
The $SNR$ is customary given in dB and it is obtained by dividing
the output signal power level $P_1$ to the noise level signal
$N_1$. Both quantities are measured at the frequency
of the modulating signal.

We investigate both the effect of varying the amplitude and 
the frequency of the modulating signal on $\eta$ and $SNR$. 
We first consider the role of the frequency of the modulating 
signal. Specifically we investigate the SR phenomenon as a 
function of $V_{rms}$ by keeping $V_s$ constant (we choose 
$V_s=0.067$ V) and by varying $f_s$ from 1 to $10^6$ Hz. 
For each pair of the control parameters $V_s$ and $f_s$,
we vary $V_{rms}$ from 0.67 to 5.33 V. The measured 
values of the power amplification $\eta$ are collected in Fig. 4,
where we show $\eta$ as a function of $V_{rms}$ for 7 different
values of $f_s$, which are 1, 10, 100, $10^3$, $10^4$, $10^5$ and 
$10^6$ Hz. The classical profile of the SR phenomenon \cite{Gammaitoni98}
is observed for the lowest values of $f_s$. For higher 
values of $f_s$, $\eta$ deviates from the canonical SR profile
by lowering and broadening its maximum. These results are in 
qualitative agreement with the explicit results theoretically 
obtained for the signal power amplification in a model bistable 
system \cite{Jung91}.   

The next investigation of the power amplification $\eta$ concerns
the study performed by keeping $f_s$ constant whereas $V_s$ is
varied. We set $f_s=10$ Hz and we vary $V_s$ from 0.0067 
to 1.00 V. For all the selected values of $V_s$ we check that the 
amount of the amplitude of the modulating signal is not sufficient to induce 
deterministic jumps between the two wells. In Fig. 5 we show the 
experimental values of $\eta$ obtained for 8 different values of
$V_s$. The selected values are 0.0067, 0.017, 0.033, 0.067, 0.167, 0.333,
0.667 and 1.00 V. In the figure the top line corresponds
to $V_s=0.0067$ V whereas the bottom line refers to the value
$V_s=1.00$ V. 
The profile of $\eta$ becomes progressively more sharp
around the value $V_{rms} \approx 1.5$ V for decreasing values 
of $V_s$. This experimental finding is in qualitative agreement 
with the results of theoretical calculations of Ref. \cite{Jung91}. 
In the figure, the curves measured for highest values
of $V_s$ tend to collapse into a unique curve that the theory indicates
as the limit behavior predicted by the linear response theory for 
negligible values of the amplitude of the modulating signal.
On the other hand, a difference is detected 
when one considers the lowest values of $V_s$. In these cases we 
experimentally detect a deviation from the expected limit curve.
These deviations are observed in our experiment because for these
values of $V_s$ (0.0067, 0.017 and 0.033 V) the signal $P_1$ becomes
of the same level of the noise level $N_1$ and it is therefore
indistinguishable from it. One can 
verify quickly the above statements by inspecting Fig. 6 where
we present the $SNR$ measured under the same conditions of Fig. 5.
In Fig. 6 the bottom curve refers to the case $V_s=0.0067$ V whereas 
the top curve is obtained by setting $V_s=1.00$ V. From the figure
it is evident that for $V_s$ equal to 0.0067, 0.017 and 0.033 V, the
$SNR$ becomes zero within the experimental errors for a wide range 
of values of $V_{rms}$. 
This effect is reflected into the deviation
of $\eta$ from the limit curve observed in Fig. 5. 

A simultaneous inspection of Figs 5 and 6 show that the results
obtained with the highest values of $V_s$ are associated with 
high values of the $SNR$ but are at the same time seriously affected 
by nonlinear distortion. This nonlinear distortion manifests 
itself in the broadening of the SNR curve. In other words, 
by using high values of $V_s$ it is possible to detect a wide 
interval of $V_{rms}$ where the SR phenomenon occurs, however 
this interval is not well described 
in terms of linear response theory. On the other hand by using 
low values of
$V_s$ one observe experimentally SR on a more limited interval of 
$V_{rms}$ but the experimental results are in this interval 
well described by a linear response theory. Hence from an experimental
point of view the more straightforward investigation of the SR phenomenon
requires the selection of a value of the amplitude of the
modulating signal which allows the detection of a large but
undistorted signal. In our case this condition is attained
when $V_s \approx 0.067$ V.

We also investigate the detailed behavior of the output noise 
power level $N_1$ at the frequency of the
modulating signal. In Fig. 7 we show $N_1$ as a
function of $V_{rms}$ for several values of $V_s$
ranging from $0.0067$ to $1.00$ V. 
In these investigation 
$f_s$ is kept constant at the value of 10 Hz.
We observe that the noise power $N_1$ slightly decreases
in an interval of values of $V_{rms}$ by increasing $V_s$. 
The noise level $N_1$ sharply increases
at the onset of the SR phenomenon, reaches a maximum and then
decreases. Depending on the value of $V_s$, the noise level may 
decrease monotonically to the asymptotic value observed for 
high values of $V_{rms}$ or reach a minimum value and then
increases until reaching the same asymptotic value. In other words,
we detect a dip in the noise level for a finite value of 
$V_{rms}$ for high values of $V_s$.  
The dip is shown in the inset of Fig. 7 for the measurements
done by setting $f_s=10$ Hz. The noise dip is more pronounced 
for high values of the signal amplitude.
A similar behavior is also observed  
for values of $f_s$ satisfying the condition $f_s < 1$ kHz.
 By taking into account the results previously 
obtained concerning the deviation form the behavior predicted 
in terms of linear response theory, we conclude that 
this behavior is belonging to the nonlinear response regime 
of stochastic resonance. In this regime, sometimes called 
weak-noise limit, the amplitude of the periodic signal can 
not be regarded as weak with respect to the noise intensity 
($V_s \gg V_{rms}$) and the linear-response theory or the 
perturbation theory is no longer valid \cite{Gammaitoni98,Shneid,Stocks}.

In the nonlinear response regime we experimentally detect 
the phenomenon of phase and frequency locking. Specifically
by increasing the value of the amplitude of the modulating signal
one observes jumps between the two stable states occurring
at phases which are progressively more synchronized with 
the phases of the modulating signal. Moreover a locking between 
the period of the output signal and the period of the input signal
is also observed for given values of the noise amplitude. 
We address this last phenomenon as frequency
locking. An example of phases and frequency locking is shown 
in Fig. 8 where we show the digitized time series of $v_d (t)$
and $v_s (t)$ recorded by setting $f_s=10$ Hz, $V_s=0.667$ V and
$V_{rms}=1.33, 2.00$ and $4.67$ V for the top, middle up and 
middle down time series of Fig. 8. The bottom time series of 
Fig. 8 shows the time evolution of the modulating signal.
All the time series are digitized synchronously. By inspecting
Fig. 8 one notes that for low levels of noise amplitude 
($V_{rms}=1.33$ V, top time series) the system jumps randomly 
from one state to the other but the jumps are statistically 
synchronized in phases. In fact they occur preferentially 
at times $t=n T/2$, where $T$ is the period of the modulating
signal and $n$ is an integer. When the noise
amplitude is increased ($V_{rms}=2.00$ V, middle top time series)
we still observe a phase synchronization, but in this case 
jumps occurs preferentially at times $t=n T/2+T/4$. Moreover 
for the present value of the noise amplitude jumps occurs with
probability almost one at each period. This means that in addition
to the phase synchronization we also observe 
frequency synchronization.  
It is worth pointing out that the noise amplitude at which we
observe phase and frequency synchronization is always detected 
in the region of the dip observed in the noise level of Fig. 7.
In other words the dip of the noise may be interpreted as 
a manifestation of the fact that phases and frequency locking 
are simultaneously present.
By increasing the noise amplitude ($V_{rms}=4.67$ V, 
middle bottom time series) jumps becomes very frequent inside 
a single period of the modulation signal so that phase
and frequency synchronization is progressively lost.
A similar effect has been theoretically considered 
in the literature \cite{Freund99} recently.

\section{Conclusions}

We report an experimental study of stochastic resonance in a physical
system. Our physical system, which is characterized by versatility
and high stability, allows us to investigate with high 
precision the SR phenomenon in a wide range of parameters,
such as the frequency and the amplitude of the
modulating signal and the noise amplitude.
In the experiments presented here, the frequency range is 
spanning up to seven orders of magnitude whereas the amplitude range 
spans  more than two orders of magnitude.

Theoretical and experimental investigations have been mainly 
focused on the linear regime of SR. However for a complete 
description of the SR phenomenon it is also important to investigate 
the nonlinear response regime of SR. We experimentally investigate 
the degree of consistence
of our experimental results with the results expected in terms
of the linear response theory. We find a range of experimental
parameters within which the linear response theory describes quite 
well the investigated dynamics. However, outside these intervals, 
nonlinear deviations from the prediction of the linear response
theory are clearly detected. These deviations primarily 
manifest them-self (i) in a saturation of the output power 
spectral density signal and of the signal 
amplification and (ii) in a non-monotonic behavior of the output
noise level associated with a high degree of phase and frequency 
synchronization.

We wish to thank ASI, INFM and MURST for financial support.

\begin{figure}
\epsfxsize=3.0in
\epsfbox{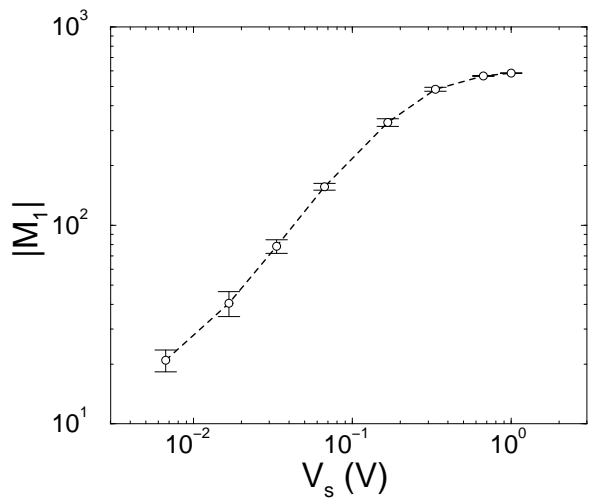}
\caption{ The amplitude $|M_1|$ of the output signal $v_d(t)$
at the frequency of the modulating signal as a function of the 
amplitude of the modulating signal $V_s$. The noise level
$V_{rms}$ and the frequency of the modulating signal $f_s$
are kept constant at the values $V_{rms}=1.89$ V and 
$f_s=10$ Hz. A linear relation between $|M_1|$ and $V_s$ 
is detected within the interval $0.017 \le V_s \le 0.067$.}
\label{fig1}
\end{figure}

\begin{figure}
\epsfxsize=3.0in
\epsfbox{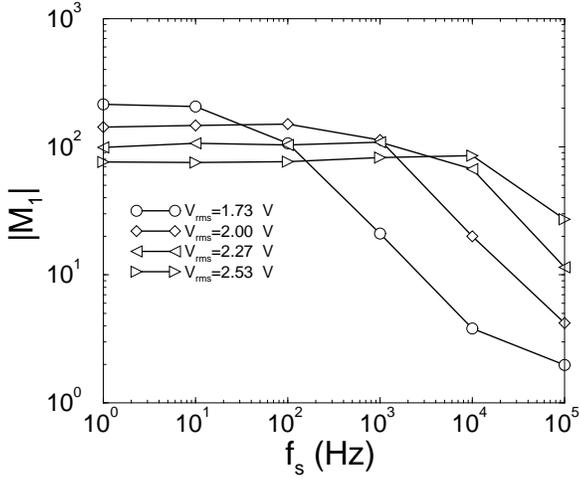}
\caption{ The amplitude $|M_1|$ of the output signal $v_d(t)$
at the frequency of the modulating signal as a function of the 
frequency of the modulating signal $f_s$. The amplitude of 
the modulating signal $V_s$ is kept constant at the value 
$V_{s}=0.067$ V whereas the noise level takes four
different values: 1.73, 2.00, 2.27 and 2.53 V. 
For each value of the noise amplitude $|M_1|$ presents
two regimes. An almost constant regime for 
$f<<\lambda_{min}/2\pi$ and a power-law decreasing regime 
(with exponent approximately -1.3) for $f>>\lambda_{min}/2\pi$.
Experiments clearly show that $\lambda_{min}$ is controlled 
by the value of $V_{rms}$.}
\label{fig2}
\end{figure}

\begin{figure}
\epsfxsize=3.0in
\epsfbox{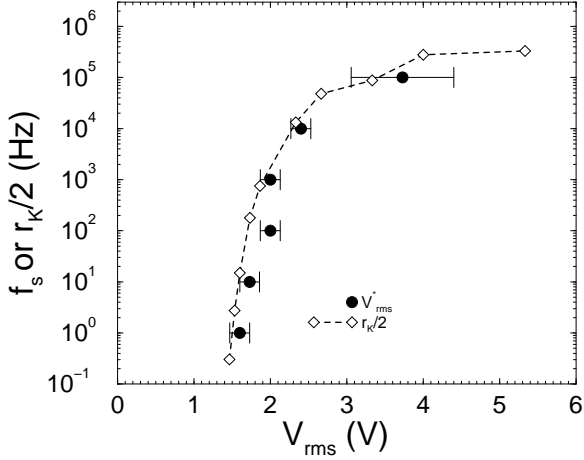}
\caption{ Comparison of (i) the occurrence of the maximal
value of the $SNR$ (black circles) as a function of 
$V_{rms}$ and $f_s$ (error bars indicate the experimental
uncertainty in the determination of the maximal $SNR$ 
with (ii) the experimental values of half of the Kramers rate
$r_k/2$ observed in the same system in the absence of the 
modulating signal. In all the SR measurements
the amplitude of the modulating signal is set to 0.067 V.
The stochastic synchronization at the SR between half 
of the Kramers rate and the frequency of the modulating 
signal is experimentally observed in a frequency interval 
spanning 6 orders of magnitude}
\label{fig3}
\end{figure}

\begin{figure}
\epsfxsize=3.0in
\epsfbox{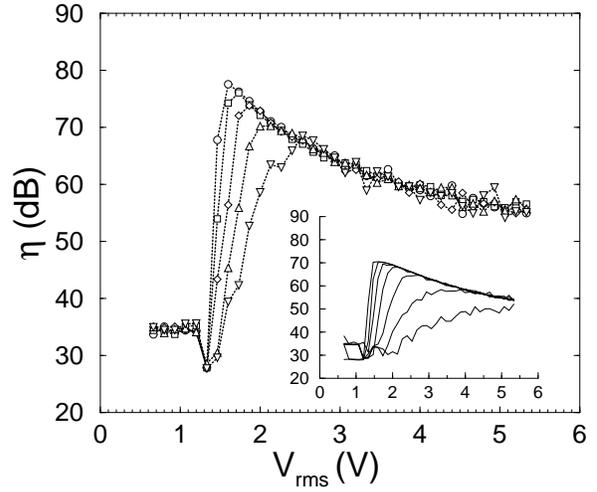}
\caption{The amplitude $|M_1|$ of the output signal $v_d(t)$
at the frequency of the modulating signal as a function of the 
noise amplitude. The amplitude of 
the modulating signal $V_s$ is kept constant at the value 
$V_{s}=0.067$ V whereas the frequency takes five
different values: 1 (circles), 10 (squares), 100 
(diamonds), 1,000 (triangles up) and 10,000 (triangles 
down) Hz. The profile of $|M_1|$ becomes progressively 
more sharp as the frequency decreases. In the inset we show 
the result of the same kind of measurements performed in 
the nonlinear regime ($V_s=0.33$ V). In this case, higher 
values of the modulating frequency are investigated. 
Specifically the curves shown refer to values  1, 
10, 100, 1,000, 10,000, 100,000 and 1,000,000 Hz,
from top to bottom respectively. Also in this case
the profile of $|M_1|$ becomes progressively 
more sharp as the frequency decreases but the low frequency 
limit is less sharp than expected from the linear response
theory.}
\label{fig4}
\end{figure}

\newpage

\begin{figure}
\epsfxsize=3.0in
\epsfbox{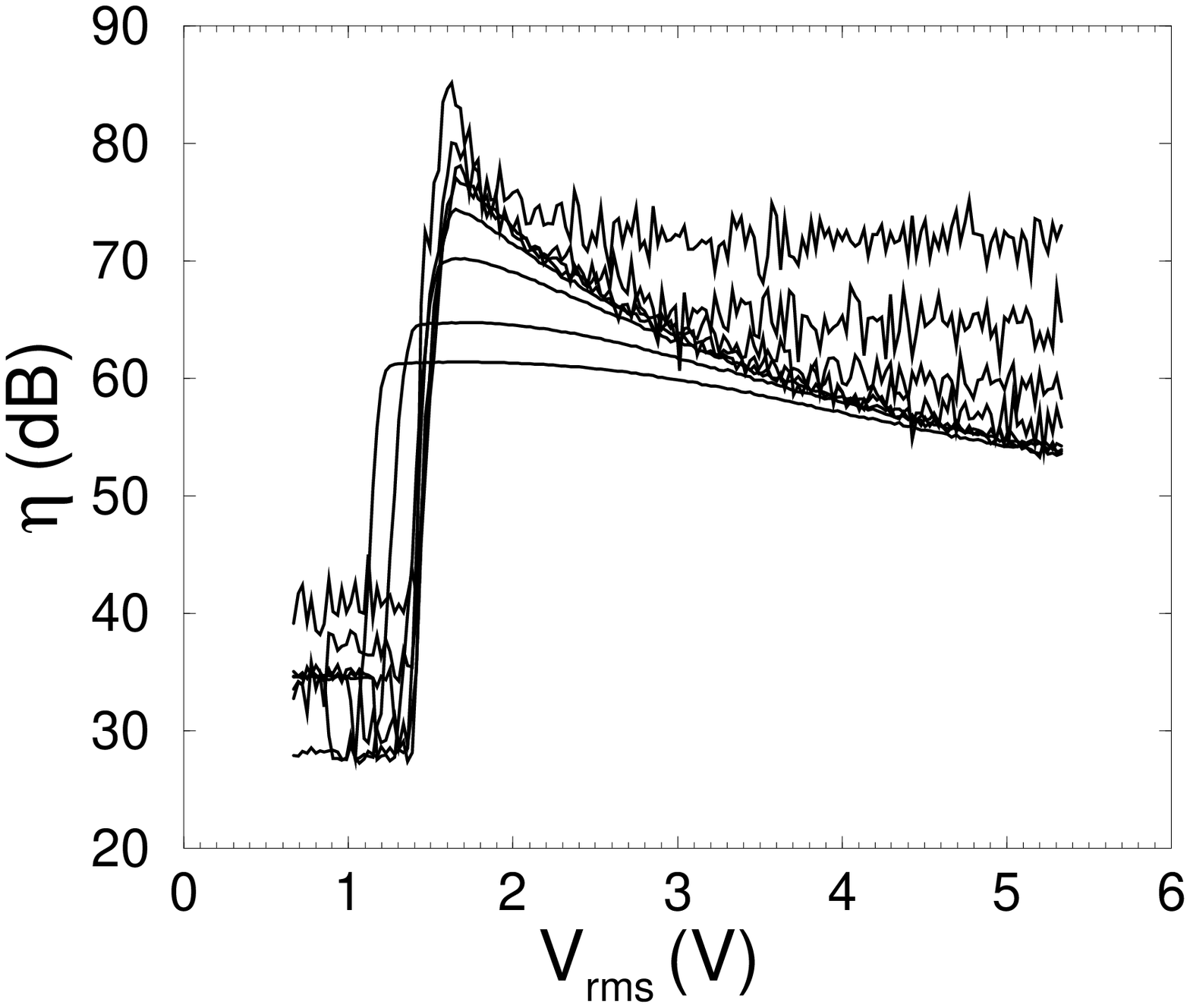}
\narrowtext
\caption{The amplitude $|M_1|$ of the output signal $v_d(t)$
at the frequency of the modulating signal as a function of the 
noise amplitude. The frequency of 
the modulating signal $f_s$ is kept constant at the value 
$f_{s}=10$ Hz whereas the amplitude takes eight 
different values 0.0067, 0.017, 0.033, 0.067, 0.167,
0.333, 0.667 and 1.00 V, which correspond to the lines 
shown from top to bottom respectively.
By diminishing the amplitude $V_s$ the amplitude $|M_1|$
progressively approaches the limit predicted by the 
linear response theory. The convergence breaks down 
for low values of $V_s$ because for such a low level of 
$|M_1|$ the noise mask the presence of the signal.
Hence the best approximation of a system evolving in the
regime well described by the linear response theory is
observed when $V_s=0.067$ V (fourth line starting from top).}
\label{fig5}
\end{figure}

\begin{figure}
\epsfxsize=3.0in
\epsfbox{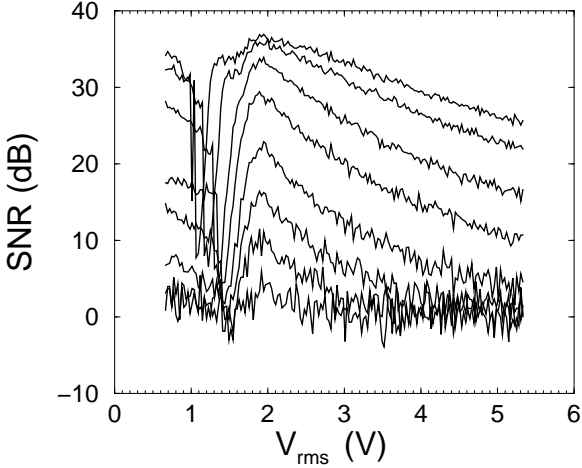}
\caption{Signal to Noise ratio (SNR) of the output signal 
$v_d(t)$ at the frequency of the modulating signal as a 
function of the noise amplitude. The frequency of 
the modulating signal $f_s$ is kept constant at the value 
$f_{s}=10$ Hz whereas the amplitude $V_s$ takes eight 
different values 0.0067, 0.017, 0.033, 0.067, 0.167,
0.333, 0.667 and 1.00 V, which correspond to the lines 
shown from bottom to top respectively.
The customary general profile of the stochastic resonance 
is observed. However shape differences are observed 
by investigating the phenomenon at different values 
of $V_s$. Specifically, high values of $V_s$ (top curves) 
are characterized by distortions introduced by the 
active nonlinearities whereas at low values of $V_s$
(bottom curves) the SNR becomes negligible. As expressed 
in the caption of Fig. 5, the experimental condition
better interpreted in terms of the linear response theory
is the one observed for $V_s=0.067$ V (fourth line from
bottom).}
\label{fig6}
\end{figure}

\begin{figure}
\epsfxsize=3.0in
\epsfbox{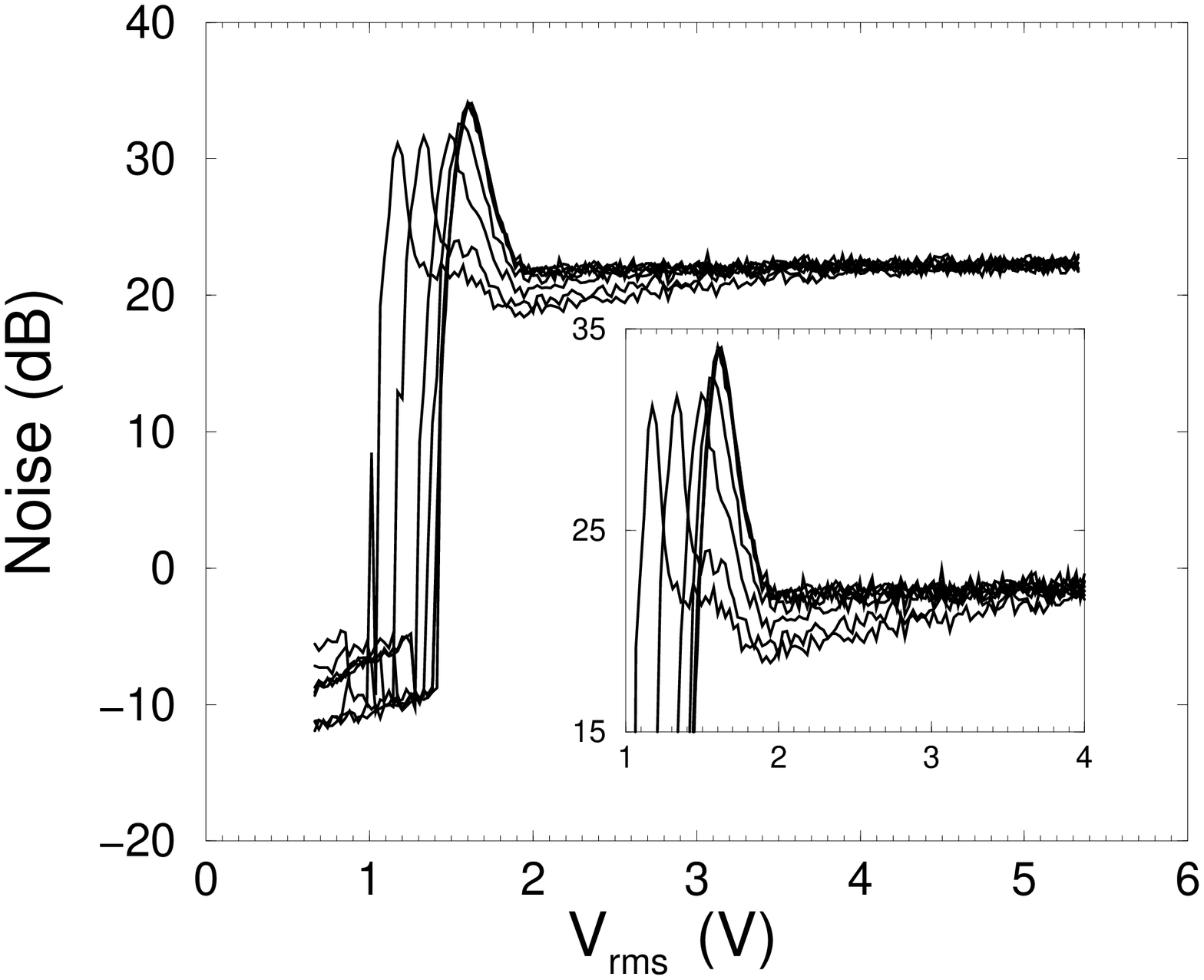}
\caption{Noise level present in the output signal $v_d(t)$
at the frequency of the modulating signal as a function of the 
noise amplitude. The frequency of 
the modulating signal $f_s$ is kept constant at the value 
$f_{s}=10$ Hz whereas the amplitude $V_s$ takes eight 
different values 0.0067, 0.017, 0.033, 0.067, 0.167,
0.333, 0.667 and 1.00 V, which correspond to the lines 
shown from top to bottom respectively.
By diminishing the amplitude $V_s$ the amplitude $|M_1|$
progressively approaches the curve observed in the
absence of modulating signal (indeed the four curves 
with lowest values of $V_s$ are almost indistinguishable 
from the $V_s=0$ V observation). For the highest
values of $V_s$ the presence of a dip is observed.
The dip is more pronounced for higher value of $V_s$.
When $V_s=0.667$ and $V_s=1.00$ V detailed structures 
emerge in the vicinity of the dip (see the inset for a 
blow-up of the region). These structures are
responsible for the structure observed in the SNR
near the maximum for highest values of $V_s$.}
\label{fig7}
\end{figure}

\begin{figure}
\epsfxsize=3.0in
\epsfbox{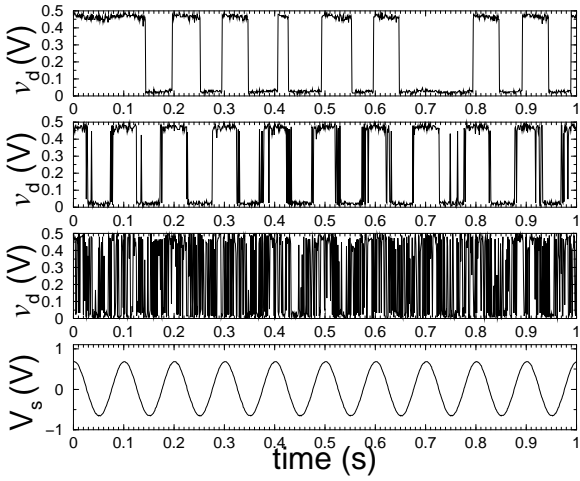}
\caption{Time evolution of $v_d(t)$ measured in the 
nonlinear regime for three different values of the
noise amplitude $V_{rms}$. All the time evolutions 
(top $V_{rms}=1.33$ V, middle up $V_{rms}=2.00$ V and 
middle down $V_{rms}=4.67$ V graph) are
synchronously recorded with respect to
the modulating signal (shown at the bottom of the figure).}
\label{fig8}
\end{figure}

\end{document}